\documentclass[conference,letterpaper]{IEEEtran}
\usepackage{graphicx}
\usepackage{amssymb}
\usepackage{epstopdf}
\usepackage{physics}
\usepackage{pst-optexp}
\usepackage{mathtools}
\usepackage{tikz}
\usepackage{subcaption}
\usepackage{xcolor}
\usepackage[hidelinks]{hyperref}

\def\editcolor{black}

\usepackage{stfloats}

\usetikzlibrary{positioning,arrows.meta}
\usepackage[utf8]{inputenc} 
\usepackage[T1]{fontenc}
\usepackage{url}
\usepackage{ifthen}
\usepackage{cite}

\begin{document}
  
\title{A novel multi-photon entangled state with enhanced resilience to path loss}

\author{%
   \IEEEauthorblockN{Armanpreet Pannu\IEEEauthorrefmark{1}\IEEEauthorrefmark{2},
                     Amr S. Helmy\IEEEauthorrefmark{1},
                     and Hesham El Gamal\IEEEauthorrefmark{2}}
    \IEEEauthorblockA{\IEEEauthorrefmark{1}%
   The Edward S. Rogers Department of Electrical and Computer Engineering, University of Toronto}
   \IEEEauthorblockA{\IEEEauthorrefmark{2}%
   School of Electrical and Information Engineering, Faculty of Engineering, The University of Sydney}
 }

\maketitle


\begin{abstract}
In the realm of quantum information, entanglement stands as a cornerstone phenomenon. It underpins a vast array of quantum information processes, offering significant potential for advancements in quantum computing, communication, and sensing. This paper introduces a novel multi-photon entangled state, which generalizes the maximally entangled single-photon state and exhibits remarkable resilience to signal attenuation in photonic applications. We demonstrate the novelty of the proposed state through a simplified target detection model and illustrate its superior performance over traditional single-photon protocols, attributed to its higher entanglement level and enhanced noise suppression capabilities. Our findings suggest that the proposed multi-photon state holds significant promise for enhancing the efficiency and reliability of photonic applications subject to loss. This work lays the groundwork for future exploration into the practical applications of multi-photon entangled states in quantum technologies, potentially revolutionizing our approach to quantum sensing and beyond.
\end{abstract}

\section{Introduction}
Described by Einstein as "spooky action at a distance," entanglement represents a unique and counter-intuitive aspect of quantum mechanics, playing a crucial role in quantum information. Two quantum systems are considered entangled if their respective quantum states cannot be described independently. In such cases, the systems exhibit non-classical correlations, even when separated by significant distances \cite{bells_thm_original_paper}. This phenomenon can be harnessed in various engineering applications, including quantum computing \cite{feynman_orignial_simulating}, quantum communication and key distribution \cite{Bennett_original_qkd}, quantum sensing \cite{lloyd_original_QI}, among others \cite{book_entaglement_engineering_and_applications}.

A class of quantum states of particular interest for their entanglement properties are the high-dimensional Bell states \cite{bell_inequality_arbitraily_high_dim}. In this paper, we look at the simplest such two-partite, $M$-qudit state:
\begin{align}
\ket{\psi_1} = \sum_{k=1}^M \ket{k}_A\ket{k}_B
\end{align}
where $A$ and $B$ represent two distinct quantum systems, and $\ket{k}$ denotes the system being in the $k$th state (out of $M$ possible states). The case of $M=2$ corresponds to one of the four Bell states, which are the maximally entangled qubit states. This state's entanglement level increases with an increase in the dimension of each Hilbert space $M$ \cite{bell_inequality_arbitraily_high_dim} \cite{violations_of_local_realism_by_Ndit} and it has been studied for enhancing channel capacity in communications \cite{beating_channel_capacity_with_etangled_ququarts}, improving noise resilience in target detection and LIDAR \cite{lloyd_original_QI}, and for quantum key distribution \cite{security_QKD_using_qudits}. In photonics, this state is typically generated by entangling two photons across $M$ modes, including orbital angular momentum modes \cite{entaglement_of_orbital_angular_momentum_photons}, time-bin modes \cite{time_bin_entagled_qubit_by_fs_pulses}, frequency modes \cite{generating_distributing_frequency_entagled_qudits}, or even spatial modes \cite{high_dim_spatial_entanglement_photons}.

A notable challenge with using the state $\ket{\psi_1}$ in photonic applications is its nature as a single photon state. Many applications involve transmitting one set of spatial modes through a lossy channel, where the single photon is either transmitted or lost entirely. This contrasts with classical protocols where multiple photons are transmitted simultaneously, and the attenuated light suffices to achieve the desired result. While this issue can be mitigated by sending multiple copies of $\ket{\psi_1}$, it either requires multiple sequential runs, consuming considerable time, or the use of many sets of $M$ orthogonal modes, which could instead be used to increase entanglement.

Most studies on multi-photon entangled states, often involve NOON states, where all photons occupy the same mode \cite{phase_measurment_multiphoton_entangled_state}, or the use of additional orthogonal modes without directly taking multiple copies of a single entangled photon \cite{multi_photon_entanglement_in_high_dimensions}. With NOON states, losing a single photon due to attenuation (mathematically equivalent to applying an annihilation operator) disrupts all entanglement between the systems, while using additional orthogonal again waste modes which could be used for increased entanglement.

Quantum sensing states that are resilient to loss have been studied and shown to exist \cite{robust_quantum_metrology} and are of great interest for both sensing and fault-tolerant quantum communication. In this paper, we propose and explore the theoretical properties of a novel multi-photon generalization of the state $\ket{\psi_1}$, employing $M$ orthogonal modes and retaining the entanglement properties of any returning photons, even if many photons are absorbed. Utilizing this state enables leveraging some of the entanglement properties of the high-dimensional Bell state, even in scenarios where the transmitted signal may be heavily attenuated.

\section{Multi-Photon entangled state}

Before diving into the core of our discussion, we establish some notation to pave the way for a smoother understanding. In dealing with $M$ modes of photons, we use the notation $\ket{\mathbf{n}}$ to represent the photon number basis state, where the boldface $\mathbf{n}$ signifies a vector. For $\mathbf{n} = (n_1, ..., n_M)$, the state $\ket{\mathbf{n}}$  represents a state with $n_i$ photons in mode $i$. {\color{\editcolor}The operators $\hat{a}^\dagger_i$ and $\hat{a}_i$ will represent the quantum creation and annihilation operator respectively for the $i$-th mode. These operators act on the photon number states as follows: $\hat{a}_i^\dagger\ket{n_1,...,n_i,...,n_M}=\sqrt{n_i+1}\ket{n_1,...,n_i+1,...,n_M}$ and $\hat{a}_i\ket{n_1,...,n_i,...,n_M}=\sqrt{n_i}\ket{n_1,...,n_i-1,...,n_M}$}. Whenever we mention a vector $\mathbf{n}$, the corresponding non-bold variable will represent a sum of the elements $n = |\mathbf{n}| = n_1 + n_2 + ... + n_M$ (This is the $l_1$ norm when all elements are positive as is the case in this paper). Extending this to multiple spatial modes, each with $M$ modes, we use $\ket{\mathbf{n},\mathbf{m}}$ to denote the basis states for two spatial modes, where $\mathbf{n}$ and $\mathbf{m}$ are photon numbers for each set of spatial modes.

With this in mind, we consider the $N$ photon state across two sets of $M$ modes which we refer to as the signal and the idler:
\begin{align}
    \ket{\psi_N}&= \frac{1}{\sqrt{C_N}}  \Bigg(\sum_{i=1}^M \hat{a}_{I_i}^\dagger\hat{a}_{S_i}^\dagger \Bigg)^N \ket{vac}  \label{eq:psi_n_operator_rep}
    \\&= 
    {N+M-1 \choose N}^{-\frac{1}{2}}\sum_{|\mathbf{N_S}|=N}\ket{\mathbf{N_S},\mathbf{N_S}} \label{eq:psi_n_number_state_rep}
 \end{align}

 Here, {\color{\editcolor}$C_N$ is a normalization constant (refer to appendix A) and} $\hat{a}_{I_i}^\dagger$ and $\hat{a}_{S_i}^\dagger$ are the creation operators for the $i$-th idler and signal modes. The sum in (\ref{eq:psi_n_number_state_rep}) extends over all vectors $\mathbf{N_S}$ with total photon number $N$. The number of such vectors is equal to ${N+M-1 \choose N}$ which gives the normalization constant in (\ref{eq:psi_n_number_state_rep}). In the case of $N=0$, we simply have the vacuum state while the case of $N=1$ gives us the high-dimensional Bell state $\ket{\psi_1}$.

We begin by noting that $\ket{\psi_N}$ spans a larger Hilbert space than $\ket{\psi_1}$, suggesting a potential for higher entanglement gains. However, these states become particularly intriguing when we explore the effects of attenuation. To delve into their distinctive properties, we introduce the operator $\hat{A}^\dagger = \sum_{i=1}^M \hat{a}_{S_i}^\dagger \hat{a}_{I_i}^\dagger$. Employing this operator enables the recursive generation of the states $\ket{\psi_N}$ via the relation:
\begin{align}
    \sqrt{N}\sqrt{N+M-1} \ket{\psi_N}&= 
    \hat{A}^\dagger\ket{\psi_{N-1}}
 \end{align}
 With a straightforward computation, we can also show that the following commutation relations hold:
 \begin{align}
    \left[\hat{a}_{S_j},\hat{A}^\dagger\right]=\hat{a}_{I_j}^\dagger 
    \qquad\qquad
    \left[\hat{a}_{I_j},\hat{A}^\dagger\right]=\hat{a}_{S_j}^\dagger
 \end{align}
 Considering the scenario where a photon from the signal data is lost, modeled by the annihilation operator $\hat{a}_{S_j}$, we can use the recursive relationship and the commutator to show by induction that the state evolves as follows:
 \begin{align}
    \hat{a}_{S_j}\ket{\psi_N}&=
    \frac{\sqrt{N}}{\sqrt{N+M-1}}\hat{a}_{I_j}^\dagger\ket{\psi_{N-1}} \label{eq:annihilate_from_psi_N}
 \end{align}

\newcounter{storeeqcounter}
\newcounter{tempeqcounter}
\addtocounter{equation}{1}%
\setcounter{storeeqcounter}%
 {\value{equation}}%
\begin{figure*}[b!]
    \normalsize
    \setcounter{tempeqcounter}{\value{equation}} 
    \begin{IEEEeqnarray}{rCl}
  \setcounter{equation}{\value{storeeqcounter}} 
  \ket{\psi_N}_{I,S} \otimes \ket{\mathbf{0}}_B &&= {N+M-1 \choose M-1}^{-\frac{1}{2}}\sum_{|\mathbf{N_S}|=N}\prod_{i=1}^M \frac{1}{\sqrt{N_{S_i}}!} (\hat{a}_{S_i}^\dagger)^{N_{S_i}} \ket{\mathbf{N_S},\mathbf{0},\mathbf{0}}_{I,S,B}
  \label{eq:floatingeq_first}
  \\&&\rightarrow
  {N+M-1 \choose M-1}^{-\frac{1}{2}}\sum_{|\mathbf{N_S}|=N}\prod_{i=1}^M \frac{1}{\sqrt{N_{S_i}}!} (\sqrt{\eta}\hat{a}_{S_i}^\dagger + \sqrt{1-\eta}\hat{a}_{B_i}^\dagger )^{N_{S_i}} \ket{\mathbf{N_S},\mathbf{0},\mathbf{0}}_{I,S,B}
  \\&&=
  {N+M-1 \choose M-1}^{-\frac{1}{2}}\sum_{|\mathbf{N_S}|=N} \sum_{|\mathbf{N_A}|\leq N}\sqrt{\eta}^{N-N_{A}}\sqrt{1-\eta}^{N_{A}}  \Bigg(\prod_{i=1}^M\sqrt{{N_{S_i} \choose N_{A_i}}} \Bigg)\ket{\mathbf{N_S},\mathbf{N_S}-\mathbf{N_A},\mathbf{N_A}}_{I,S,B}
  \\&&=
 \sum_{|\mathbf{N_A}|\leq N}\sqrt{\eta}^{N-N_A}\sqrt{1-\eta}^{N_A}\prod_{i=1}^M\frac{1}{\sqrt{N_{A_i}!}} \left(\hat{a}_{S_i}\right)^{{N_{A_i}}} \ket{\psi_N}_{I,S}\otimes \ket{\mathbf{N_A}}_B \label{eq:floatingeq_last}
    \end{IEEEeqnarray}
    \setcounter{equation}{\value{tempeqcounter}} 
    \hrulefill
    \vspace*{4pt}
  \end{figure*}

Equation (\ref{eq:annihilate_from_psi_N}) establishes the novelty of the state $\ket{\psi_N}$. In the case where we send $N$ photons and lose a single photon, the returning $(N-1)$ photons maintain their entanglement structure but the idler has an extra photon compared to $\ket{\psi_{N-1}}$. One can interpret this with a simple thought experiment where Alice sends out the $N$ signal photons to Bob who then measures and identifies a single photon and determines the mode it occupies. By doing so, Bob collapses one of the $N$ idler photons that Alice is holding to the same mode the single signal photon was found in. The remaining $(N-1)$ photons stay entangled, unaffected by the interaction of the lost photon with Bob, who represents the environment in this scenario.

This behavior is highly atypical of photons (and bosons in general) since their indistinguishability properties prevent us from thinking of them as independent particles unless they occupy disjoint Hilbert spaces. Yet, in this scenario, the photons behave as identical yet independent particles, almost as if each photon resides in its unique Hilbert space.

In practical terms, photon loss through attenuation introduces uncertainty in the number and modes of the lost photons. The transmitted state thus becomes a statistical mix of states $\prod_{i=1}^{M}(\hat{a}_{I_i}^\dagger)^{N_{A_i}}\ket{\psi_{N-k}}$ for all combinations of $N_{A_1}, ..., N_{A_M}$, where $k = N_{A_1} + ... + N_{A_M} \leq N$. These states are mutually orthogonal, allowing for measurements of the final state to account for all possible incoming states and still exploit the entanglement present in the $N-k$ returning photons.

%

\section{Quantum Target Detection}

To demonstrate the potential of this state for sensing applications, we explore the problem of target detection, where the goal is to discern the presence or absence of a target with reflectivity $0<\eta<1$ by probing with a light source. In our model, we assume that there exists background noise that could falsely signal the target's presence. In this simplified scenario, we assume that the receiver picks up either background noise or the partially reflected signal but not both at the same time (where the attenuated version of $\ket{\psi_N}$ mixes with the noise) akin to the approach in foundational target detection work \cite{lloyd_original_QI}. This model holds practical value when the average incoming noise and average reflected photons ($MP_B$ and $\eta N$) are both much less than one or in sensing applications where the goal is security and deterring bad actors from sending false reflections.

In the case where the target is present, the loss is modeled with a beamsplitter of reflectivity $\eta$ where one input is the signal and the other one is the vacuum (excluding noise mixing). In the Heisenberg picture, the creation operator evolves as $\hat{a}_{S_i}^\dagger\rightarrow \sqrt{\eta}\hat{a}_{S_i}^\dagger + \sqrt{1-\eta}\hat{a}_{B_i}^\dagger $. Evolving the signal, idler, and background modes as in equations
\eqref{eq:floatingeq_first} to \eqref{eq:floatingeq_last}
on the bottom of the page
\pageref{eq:floatingeq_first} and tracing out the background modes, we derive the collected state in the case of target present:
\begin{align}
    \rho_{pres} = \sum_{|\mathbf{N_A}|\leq N} \ket{\tilde{\phi}_{\mathbf{N_A}}}\bra{\tilde{\phi}_{\mathbf{N_A}}} \label{eq:rho_pres}
\end{align}
where
\begin{align}
    \ket{\tilde{\phi}_{\mathbf{N_A}}} 
    &=
     \sqrt{\eta}^{N-N_A}\sqrt{1-\eta}^{N_A}\prod_{i=1}^M\frac{1}{\sqrt{N_{A_i}!}} \left(\hat{a}_{S_i}\right)^{{N_{A_i}}} \ket{\psi_N}
\end{align}
Thus the collected light comes back as a mixture of $\ket{\tilde{\phi}_{\mathbf{N_A}}} $ which is the un-normalized version of the state where $\mathbf{N_A}$ photons remain in the environment. The norm of these states $ \bra{\phi_{\mathbf{N_A}}}  \ket{\phi_{\mathbf{N_A}}} $ gives the probability of measuring $\mathbf{N_A}$ photons in the environment as evident by equation (\ref{eq:rho_pres}) and the orthogonality of the states $\ket{\phi_{\mathbf{N_A}}} $. Given the unknown number of photons that remain in the environment, our measurement must account for all possibilities. Fortunately the normalized states $\ket{\phi_{\mathbf{N_A}}} = \frac{\ket{\tilde{\phi}}}{\sqrt{\braket{\tilde{\phi}_{\mathbf{N_A}}}}}$ are orthonormal, facilitating the use of the projective measurement  $\left\{\hat{P},1-\hat{P}\right\}$ where:
\begin{align}
    \hat{P}=\sum_{|\mathbf{N_A}|\leq{N-1}} \ket{\phi_{\mathbf{N_A}}}\bra{\phi_{\mathbf{N_A}}} \label{hat_P}
\end{align}
This measurement scheme excludes scenarios where $|\mathbf{N_A}| = N$, indicating total photon absorption. The probability of a positive detection, in the presence of a target, is given by:
\begin{align}
    1-P_{MD} 
    &=
    Tr\left[\hat{P} \rho_{pres} \right]
    \\&=
    \sum_{\mathbf{N_A} \leq{N-1}} \braket{\tilde{\phi}_{\mathbf{N_A}}}
    \\&=
    1-(1-\eta)^{N}
\end{align}
\begin{figure*}[]   
    \centering
    \begin{subfigure}{0.3\textwidth}
        \centering
        \includegraphics[height=1.6in]{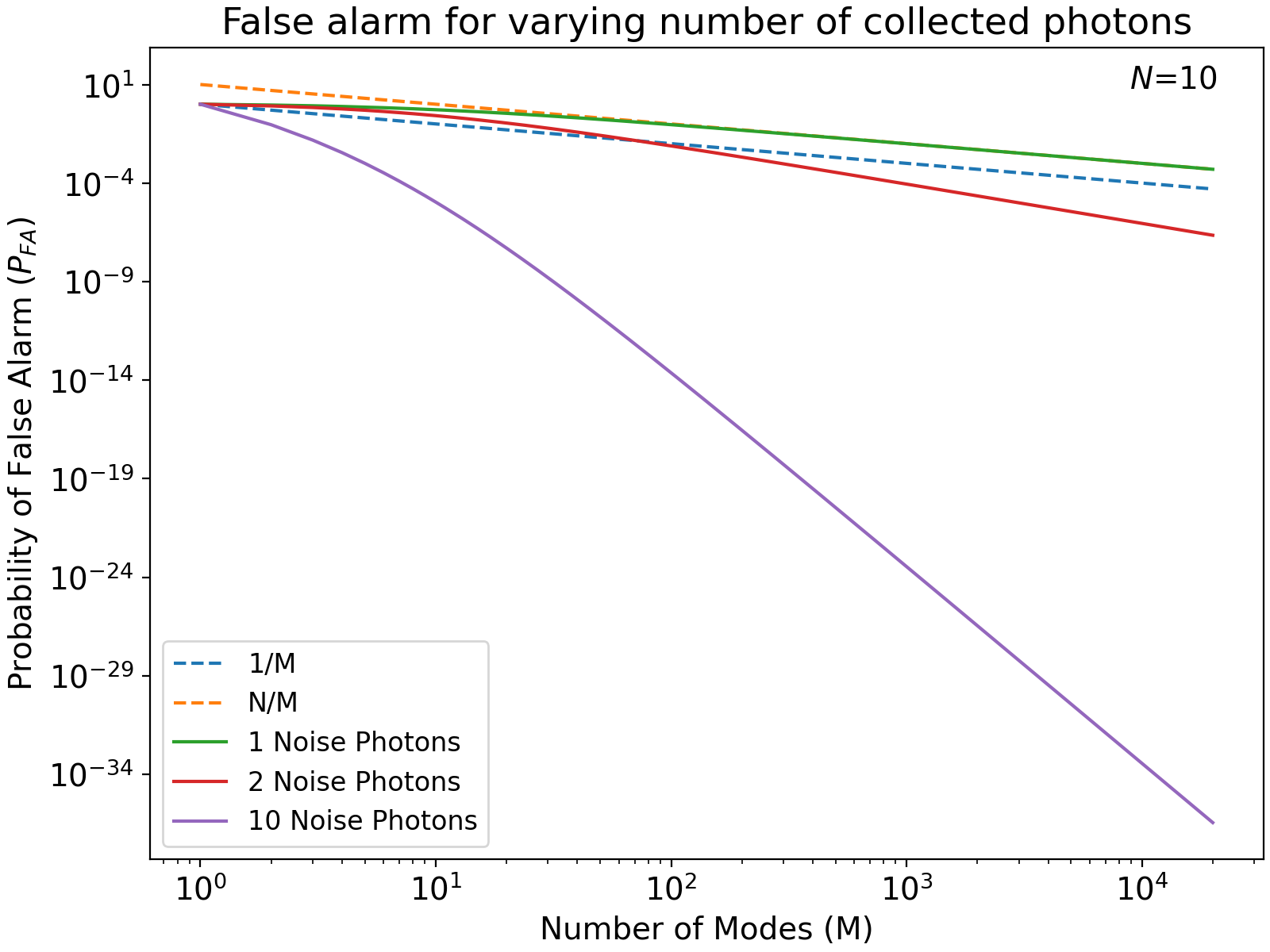}
        \caption{$N=10$}
    \end{subfigure}%
    ~ 
    \begin{subfigure}{0.3\textwidth}
        \centering
        \includegraphics[height=1.6in]{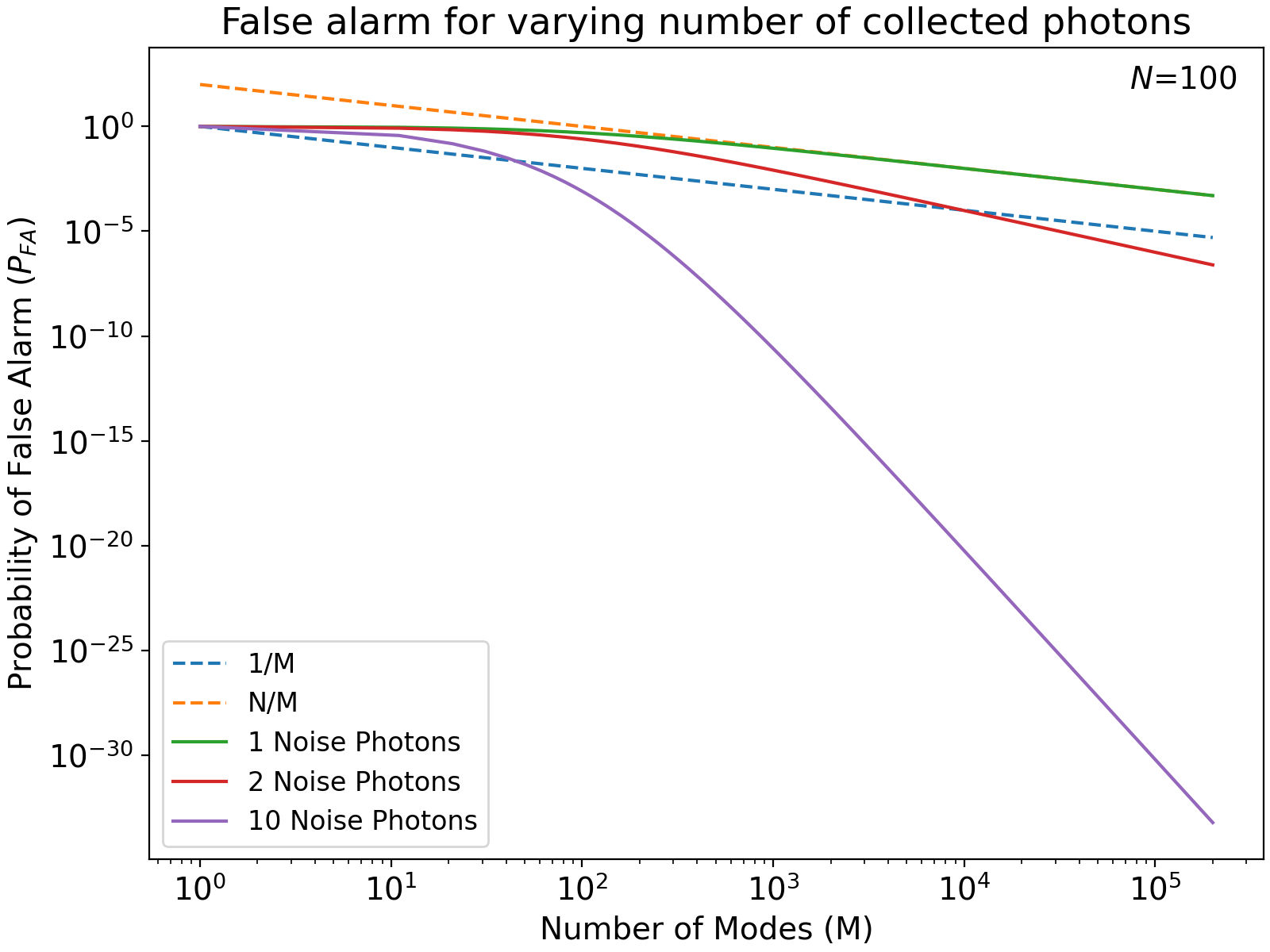}
        \caption{$N=100$}
    \end{subfigure}
    ~ 
    \begin{subfigure}{0.3\textwidth}
        \centering
        \includegraphics[height=1.6in]{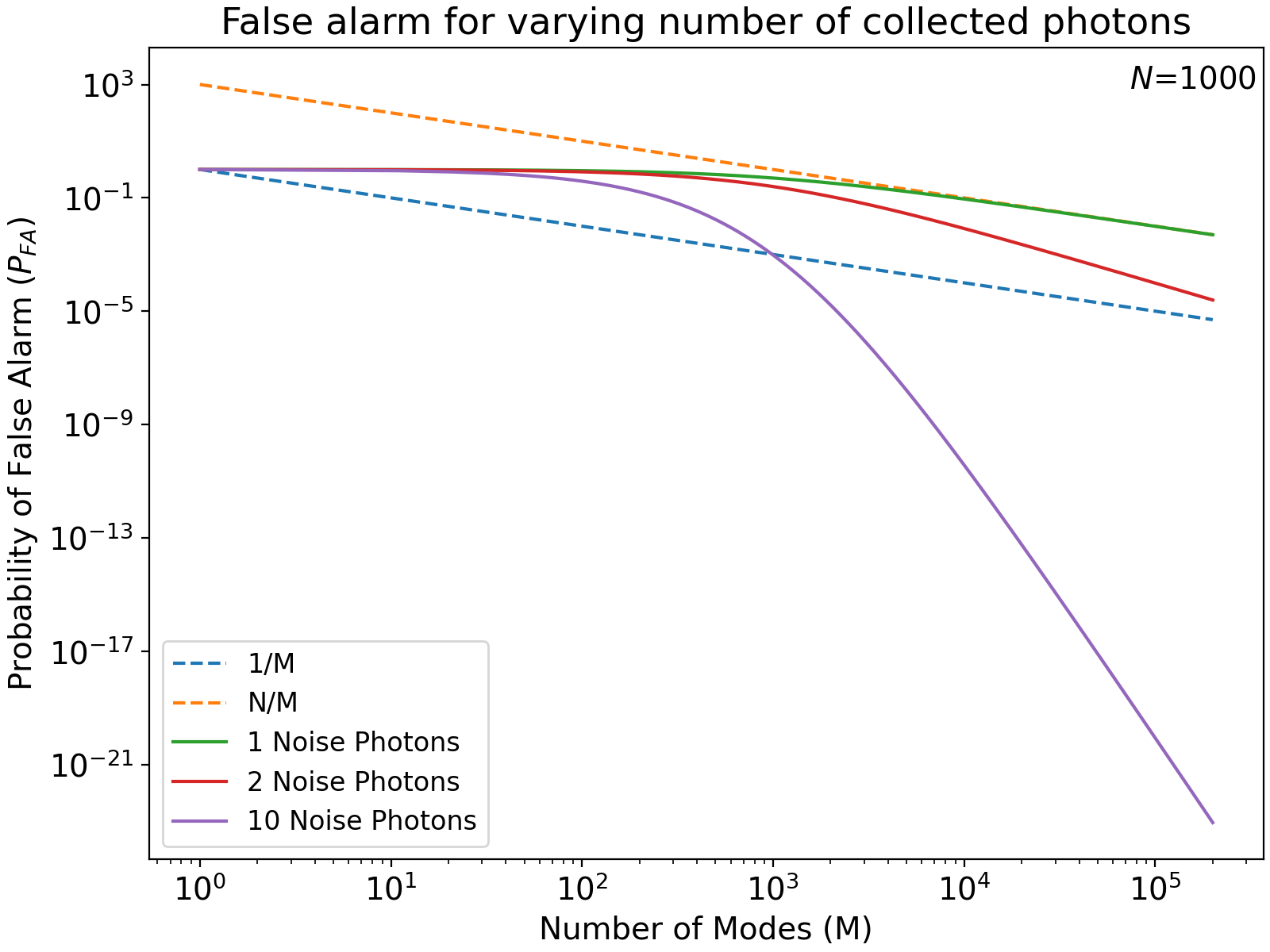}
        \caption{$N=1000$}
    \end{subfigure}
    \caption{Comparison of the Probability of false alarm for a varying number of incoming photons. The blue dashed curve is $P_{FA}=\frac{1}{M}$ which is the probability of a false alarm when one photon in the state $\ket{\psi_1}$ is transmitted. The orange dashed curve is $P_{FA}=\frac{N}{M}$ which is the $N<<M$ approximation of the probability of a false alarm if $N$ copies of the $\ket{\psi_1}$ are transmitted. The overall probability of error depends highly on how the noise is distributed.}
    \label{fig:p_fa}
\end{figure*}
This indicates that a missed detection occurs solely when all $N$ photons are absorbed, with even a single returning photon indicating target presence. However, this approach's efficacy hinges on whether utilizing $\ket{\psi_N}$ alongside $\hat{P}$ enhances noise resilience.

 In the case that the detector picks up only environmental noise, the idler mode is simply the starting state traced out and the signal is replaced with the environment. The returning state is:
 {\small
 \begin{align}
    \rho_{abs}= {N+M-1 \choose M-1}^{-1} \sum_{|\mathbf{N_S}|=N}\sum_{\mathbf{N_B}} p_{\mathbf{N_B}} \ket{\mathbf{N_S},\mathbf{N_B}}\bra{\mathbf{N_S},\mathbf{N_B}} 
\end{align} 
}
where $p_{\mathbf{N_B}}$ is the probability of having $\mathbf{N_B}$ photons in the environment. Typically, this is taken to be the probability distribution given by identical thermal states in each mode. The only assumption we make here is that this probability solely depends on the total number of noise photons so $p_{\mathbf{N_B}}=\tilde{p}_{N_B}$ (which is indeed the case for thermal noise). We highlight that this is the probability of having $N_B$ photons in a particular arrangement while the overall probability of having $N_B$ photons accounting for all arrangements is  ${N_B+M-1 \choose N_B}\tilde{p}_{N_B}$.\\
The probability of getting a false alarm can be computed as in the appendix:
\begin{align}
    P_{FA} &= \tilde{p}_{1} \frac{N}{N+M-1} + \tilde{p}_{2} \frac{N(N-1)}{(N+M-1)(N+M-2)} + \nonumber
    \\&\qquad \qquad \qquad\qquad ... + \tilde{p}_{N} {M+N-1 \choose N}^{(-1)} \label{eq:p_fa}
\end{align}
Each term in this result corresponds to a different number of photons picked up by the detector. All terms go to zero as $M \rightarrow \infty$ demonstrating that the state $\ket{\psi_N}$ has the same noise suppression property as $\ket{\psi_1}$ which is characterized by $P^{1}_{FA}\sim \frac{1}{M}$. 

In comparison to the single-photon protocol using $\ket{\psi_1}$ as in \cite{lloyd_original_QI}, the first term which corresponds to a single incoming noise photon has worse performance for all $N>1$. However, all other terms in (\ref{eq:p_fa}) eventually surpass $\frac{1}{M}$ for large enough $M$ as can be seen in figure \ref{fig:p_fa}. The last term corresponds to receiving all the $N$ transmitted photons in which case the error scales inversely to the number of terms in (\ref{eq:psi_n_number_state_rep}) and is always lower than $\frac{1}{M}$. With an increase in the number of photons returned error probabilities diminish more rapidly with $M$, leading to lower error rates due to increased entanglement.

Furthermore, if we factor in attenuation (which was the initial motivation for $\ket{\psi_N}$), a protocol using $\ket{\psi_1}$ necessitates sending $N$ copies to match the energy expenditure and expected number of reflected photons with that of a protocol using $\ket{\psi_N}$. As a result, the probability of a false alarm in the single photon case goes up to $\frac{N}{M}$, in the $N<<M$ approximation, and thus every term in (\ref{eq:p_fa}) outperforms the single-photon protocol. Intuitively, this is because transmitting a single photon across $N$ trials necessitates more frequent detector activation compared to transmitting all $N$ photons simultaneously, thus increasing noise susceptibility.

It's also important to recognize that $\tilde{p}_i$ might vary with $M$. In most practical scenarios, including those involving thermal noise, increasing $M$ tends to result in greater noise pickup. This works in our favor, as it pushes the distribution of $\tilde{p}_i$ towards higher values of $i$. This effect, in turn, shifts the false alarm probability in equation (\ref{eq:p_fa}) to the higher-order terms, which have better noise suppression.

The overall probability of error is heavily dependent on the noise profile and $\tilde{p}_i$. With a preliminary understanding of the expected photon return in the presence of a target, we can refine the projector $\hat{P}$ to focus on a narrower photon range, thereby reducing false alarm rates, especially when the anticipated number of photons due to target reflectivity, $\eta N$, is different from the expected noise photons.

\section{Conclusion}
In conclusion, we have introduced a multi-photon entangled state that extends the concept of the maximally entangled single-photon Bell state $\ket{\psi_1}$. This innovative state possesses a remarkable property: even when the signal undergoes attenuation, the entanglement structure of the returning photons is largely preserved, as though a smaller number of photons were dispatched without any loss.

Our exploration into a simplified target detection scenario highlights the unique advantages of this multi-photon state. Not only does it maintain its integrity in the face of attenuation, but it also surpasses the single-photon protocol, thanks to a higher level of entanglement and diminished noise interference due to fewer detection events.

The characteristics of the $\ket{\psi_N}$ state, as examined in this study, indicate potential benefits across a range of photonic applications that utilize the two-partite Bell state and are prone to loss. Although further investigation is necessary to determine whether $\ket{\psi_N}$ exhibits similar advantages in other contexts, particularly where noise mixing or alternative channels play a role, the increased photon count and expanded Hilbert space suggest that its susceptibility to external disturbances should be equal to or less than that of multiple copies of $\ket{\psi_1}$.

\bibliographystyle{IEEEtran}
\bibliography{bibliography}

\begin{thebibliography}{10}
\providecommand{\url}[1]{#1}
\csname url@samestyle\endcsname
\providecommand{\newblock}{\relax}
\providecommand{\bibinfo}[2]{#2}
\providecommand{\BIBentrySTDinterwordspacing}{\spaceskip=0pt\relax}
\providecommand{\BIBentryALTinterwordstretchfactor}{4}
\providecommand{\BIBentryALTinterwordspacing}{\spaceskip=\fontdimen2\font plus
\BIBentryALTinterwordstretchfactor\fontdimen3\font minus
  \fontdimen4\font\relax}
\providecommand{\BIBforeignlanguage}[2]{{%
\expandafter\ifx\csname l@#1\endcsname\relax
\typeout{** WARNING: IEEEtran.bst: No hyphenation pattern has been}%
\typeout{** loaded for the language `#1'. Using the pattern for}%
\typeout{** the default language instead.}%
\else
\language=\csname l@#1\endcsname
\fi
#2}}
\providecommand{\BIBdecl}{\relax}
\BIBdecl

\bibitem{bells_thm_original_paper}
\BIBentryALTinterwordspacing
J.~S. Bell, ``On the einstein podolsky rosen paradox,'' \emph{Physics Physique
  Fizika}, vol.~1, pp. 195--200, Nov 1964. [Online]. Available:
  \url{https://link.aps.org/doi/10.1103/PhysicsPhysiqueFizika.1.195}
\BIBentrySTDinterwordspacing

\bibitem{feynman_orignial_simulating}
R.~P. Feynman, ``Simulating physics with computers,'' \emph{International
  journal of theoretical physics}, vol.~21, no. 6/7, pp. 467--488, 1982.

\bibitem{Bennett_original_qkd}
\BIBentryALTinterwordspacing
C.~H. Bennett and G.~Brassard, ``Quantum cryptography: Public key distribution
  and coin tossing,'' \emph{Theoretical Computer Science}, vol. 560, p. 7–11,
  Dec. 2014. [Online]. Available:
  \url{http://dx.doi.org/10.1016/j.tcs.2014.05.025}
\BIBentrySTDinterwordspacing

\bibitem{lloyd_original_QI}
S.~Lloyd, ``Enhanced sensitivity of photodetection via quantum illumination,''
  \emph{Science}, vol. 321, no. 5895, pp. 1463--1465, 2008.

\bibitem{book_entaglement_engineering_and_applications}
\BIBentryALTinterwordspacing
F.~J. Duarte and T.~S. Taylor, \emph{Quantum Entanglement Engineering and
  Applications}, ser. 2053-2563.\hskip 1em plus 0.5em minus 0.4em\relax IOP
  Publishing, 2021. [Online]. Available:
  \url{https://dx.doi.org/10.1088/978-0-7503-3407-5}
\BIBentrySTDinterwordspacing

\bibitem{bell_inequality_arbitraily_high_dim}
\BIBentryALTinterwordspacing
D.~Collins, N.~Gisin, N.~Linden, S.~Massar, and S.~Popescu, ``Bell inequalities
  for arbitrarily high-dimensional systems,'' \emph{Phys. Rev. Lett.}, vol.~88,
  p. 040404, Jan 2002. [Online]. Available:
  \url{https://link.aps.org/doi/10.1103/PhysRevLett.88.040404}
\BIBentrySTDinterwordspacing

\bibitem{violations_of_local_realism_by_Ndit}
\BIBentryALTinterwordspacing
D.~Kaszlikowski, P.~Gnaciński, M.~Żukowski, W.~Miklaszewski, and
  A.~Zeilinger, ``Violations of local realism by two entangled qunits are
  stronger than for two qubits,'' \emph{Physical Review Letters}, vol.~85,
  no.~21, p. 4418–4421, Nov. 2000. [Online]. Available:
  \url{http://dx.doi.org/10.1103/PhysRevLett.85.4418}
\BIBentrySTDinterwordspacing

\bibitem{beating_channel_capacity_with_etangled_ququarts}
\BIBentryALTinterwordspacing
X.-M. Hu, Y.~Guo, B.-H. Liu, Y.-F. Huang, C.-F. Li, and G.-C. Guo, ``Beating
  the channel capacity limit for superdense coding with entangled ququarts,''
  \emph{Science Advances}, vol.~4, no.~7, p. eaat9304, 2018. [Online].
  Available: \url{https://www.science.org/doi/abs/10.1126/sciadv.aat9304}
\BIBentrySTDinterwordspacing

\bibitem{security_QKD_using_qudits}
\BIBentryALTinterwordspacing
L.~Sheridan and V.~Scarani, ``Security proof for quantum key distribution using
  qudit systems,'' \emph{Physical Review A}, vol.~82, no.~3, Sep. 2010.
  [Online]. Available: \url{http://dx.doi.org/10.1103/PhysRevA.82.030301}
\BIBentrySTDinterwordspacing

\bibitem{entaglement_of_orbital_angular_momentum_photons}
\BIBentryALTinterwordspacing
A.~Mair, A.~Vaziri, G.~Weihs, and A.~Zeilinger, ``Entanglement of the orbital
  angular momentum states of photons,'' \emph{Nature}, vol. 412, no. 6844, p.
  313–316, Jul. 2001. [Online]. Available:
  \url{http://dx.doi.org/10.1038/35085529}
\BIBentrySTDinterwordspacing

\bibitem{time_bin_entagled_qubit_by_fs_pulses}
\BIBentryALTinterwordspacing
I.~Marcikic, H.~de~Riedmatten, W.~Tittel, V.~Scarani, H.~Zbinden, and N.~Gisin,
  ``Time-bin entangled qubits for quantum communication created by femtosecond
  pulses,'' \emph{Phys. Rev. A}, vol.~66, p. 062308, Dec 2002. [Online].
  Available: \url{https://link.aps.org/doi/10.1103/PhysRevA.66.062308}
\BIBentrySTDinterwordspacing

\bibitem{generating_distributing_frequency_entagled_qudits}
R.~Jin, R.~Shimizu, M.~Fujiwara, M.~Takeoka, R.~Wakabayashi, T.~Yamashita,
  S.~Miki, H.~Terai, T.~Gerrits, and M.~Sasaki, ``Simple method of generating
  and distributing frequency-entangled qudits,'' \emph{Quantum Science and
  Technology}, vol.~1, p. 015004, 11 2016.

\bibitem{high_dim_spatial_entanglement_photons}
\BIBentryALTinterwordspacing
M.~\ifmmode~\dot{Z}\else \.{Z}\fi{}ukowski, A.~Zeilinger, and M.~A. Horne,
  ``Realizable higher-dimensional two-particle entanglements via multiport beam
  splitters,'' \emph{Phys. Rev. A}, vol.~55, pp. 2564--2579, Apr 1997.
  [Online]. Available: \url{https://link.aps.org/doi/10.1103/PhysRevA.55.2564}
\BIBentrySTDinterwordspacing

\bibitem{phase_measurment_multiphoton_entangled_state}
\BIBentryALTinterwordspacing
M.~W. Mitchell, J.~S. Lundeen, and A.~M. Steinberg, ``Super-resolving phase
  measurements with a multiphoton entangled state,'' \emph{Nature}, vol. 429,
  no. 6988, p. 161–164, May 2004. [Online]. Available:
  \url{http://dx.doi.org/10.1038/nature02493}
\BIBentrySTDinterwordspacing

\bibitem{multi_photon_entanglement_in_high_dimensions}
\BIBentryALTinterwordspacing
M.~Malik, M.~Erhard, M.~Huber, M.~Krenn, R.~Fickler, and A.~Zeilinger,
  ``Multi-photon entanglement in high dimensions,'' \emph{Nature Photonics},
  vol.~10, no.~4, p. 248–252, Feb. 2016. [Online]. Available:
  \url{http://dx.doi.org/10.1038/nphoton.2016.12}
\BIBentrySTDinterwordspacing

\bibitem{robust_quantum_metrology}
\BIBentryALTinterwordspacing
M.~Oszmaniec, R.~Augusiak, C.~Gogolin, J.~Kołodyński, A.~Acín, and
  M.~Lewenstein, ``Random bosonic states for robust quantum metrology,''
  \emph{Physical Review X}, vol.~6, no.~4, Dec. 2016. [Online]. Available:
  \url{http://dx.doi.org/10.1103/PhysRevX.6.041044}
\BIBentrySTDinterwordspacing

\end{thebibliography}

\clearpage
\onecolumn
\section{Appendix}

{\color{\editcolor}
\subsection{The states and their recursive relations} \label{appendix:the_state}
\noindent
Starting from our proposed definition (\ref{eq:psi_n_operator_rep}), we have:
\begin{align}
    \ket{\psi_N}&= \frac{1}{\sqrt{C_N}} \Bigg(\sum_{i=1}^M \hat{a}_{I_i}^\dagger\hat{a}_{S_i}^\dagger \Bigg)^N \ket{vac}
    \\&=
    \frac{1}{\sqrt{C_N}} \left(\sum_{|\mathbf{N_S}|=N}  {N \choose N_{S_1},...,N_{S_M}}\left(\mathbf{\hat{a}}^\dagger_I \mathbf{\hat{a}}^\dagger_S \right)^\mathbf{N_S}\right)\ket{vac} \label{eq:multinomial_expansion}
    \\&=
    \frac{1}{\sqrt{C_N}}\sum_{|\mathbf{N_S}|=N}N!\ket{\mathbf{N_S},\mathbf{N_S}}
 \end{align}
 Equation (\ref{eq:multinomial_expansion}) follows from the multinomial theorem where ${N \choose N_{S_1},...,N_{S_M}}=\frac{N!}{N_{S_1}!...N_{S_M}!}$ is the multinomial coefficient. The normalization constant can be computed as:
 \begin{align}
    1&=\sum_{|\mathbf{N_S}|=N}  \frac{1}{C_N} (N!)^2
    \\&=
    \frac{(N!)^2}{C_N } \sum_{|\mathbf{N_S}|=N} 1
    \\&=
    \frac{(N!)^2}{ C_N}  {{N+M-1} \choose {M-1}} \label{eq:combinatorics}
    \\C_N&= 
     \frac{N!(N+M-1)!}{(M-1)!}
 \end{align}
 Here, equation (\ref{eq:combinatorics}) follows from a standard combinatorics formula for the number of vectors with element by element sum equal to $N$. Furthermore, we have that:
 \begin{align}
    \frac{C_N}{C_{N-1}} &= N! \frac{(N+M-1)!}{(M-1)!} \frac{1}{(N-1)!} \frac{(M-1)!}{(N+M-2)!}
    \\&=
    N(N+M-1)
 \end{align}
 Which we can use to construct a recursive formula as:
 \begin{align}
    \ket{\psi_N}&= \frac{1}{\sqrt{C_N}} \Bigg(\sum_{i=1}^M \hat{a}_{I_i}^\dagger\hat{a}_{S_i}^\dagger \Bigg)^N \ket{vac}
    \\&=
    \frac{1}{\sqrt{C_N}} \Bigg(\sum_{i=1}^M \hat{a}_{I_i}^\dagger\hat{a}_{S_i}^\dagger \Bigg) \Bigg(\sum_{i=1}^M \hat{a}_{I_i}^\dagger\hat{a}_{S_i}^\dagger \Bigg)^{N-1} \ket{vac}
    \\&=
    \frac{\sqrt{C_{N-1}}}{\sqrt{C_N}} \Bigg(\sum_{i=1}^M \hat{a}_{I_i}^\dagger\hat{a}_{S_i}^\dagger \Bigg) \ket{\psi_{N-1}}
    \\&=
    \frac{1}{\sqrt{N}\sqrt{N+M-1}}
    \Bigg(\sum_{i=1}^M \hat{a}_{I_i}^\dagger\hat{a}_{S_i}^\dagger \Bigg) \ket{\psi_{N-1}}
    \\ \sqrt{N}\sqrt{N+M-1}\ket{\psi_N}&=
    \Bigg(\sum_{i=1}^M \hat{a}_{I_i}^\dagger\hat{a}_{S_i}^\dagger \Bigg) \ket{\psi_{N-1}}
    \\ \sqrt{N}\sqrt{N+M-1}\ket{\psi_N}&=
    \hat{A}^\dagger\ket{\psi_{N-1}}
 \end{align}
 where we let $\hat{A}^\dagger= \sum_{i=1}^M \hat{a}_{S_i}^\dagger \hat{a}_{I_i}^\dagger$.
 
\noindent
Now to study the behavior of the state $\ket{\psi_N}$ as a photon is removed, we first compute the commutator:
 \begin{align}
    \left[\hat{a}_{S_j},\hat{A}^\dagger\right]&= 
    \sum_{i=1}^M \left[\hat{a}_{S_j},\hat{a}_{S_i}^\dagger \hat{a}_{I_i}^\dagger\right]
    \\&=
   \left[\hat{a}_{S_j},\hat{a}_{S_j}^\dagger \hat{a}_{I_j}^\dagger\right] &
    \\&=
   \sum_{i=1}^M \Big(\hat{a}_{S_j}\hat{a}_{S_j}^\dagger \hat{a}_{I_j}^\dagger- \hat{a}_{S_j}^\dagger \hat{a}_{I_j}^\dagger\hat{a}_{S_j}\Big)
    \\
    \left[\hat{a}_{S_j},\hat{A}^\dagger\right]&=\hat{a}_{I_j}^\dagger \qquad\qquad\qquad \qquad\qquad\qquad \text{likewise:} \quad\left[\hat{a}_{I_j},\hat{A}^\dagger\right]=\hat{a}_{S_j}^\dagger
 \end{align}
 So applying an annihilation to the state, we claim that:
 \begin{align}
    \hat{a}_{S_j}\ket{\psi_N}=\frac{\sqrt{N}}{\sqrt{N+M-1}} \hat{a}_{I_j}^\dagger\ket{\psi_{N-1}} \label{eq_appendix:annihilate_from_psi_N}
 \end{align}
 We can prove this by induction on $N$. For $N=1$, this is easy to verify:
 \begin{align}
    \hat{a}_{S_j}\ket{\psi_N} &= \frac{1}{\sqrt{M}}\hat{a}_{S_j}\sum_{i=1}^M \ket{\mathbf{e}_i,\mathbf{e}_i} 
    \\&= 
    \ket{\mathbf{e}_j,0}
    \\&=
    \frac{\sqrt{1}}{\sqrt{1+M-1}} \hat{a}_{I_j}^\dagger\ket{\psi_{0}}
 \end{align}
 Now for the induction step, assume (\ref{eq_appendix:annihilate_from_psi_N}) holds for $N-1$

 \begin{align}
    \sqrt{N}\sqrt{N+M-1}\hat{a}_{S_j}\ket{\psi_N}&=
    \hat{a}_{S_j}\hat{A}^\dagger \ket{\psi_{N-1}} 
    \\&=\left(\hat{a}_{I_j}^\dagger +\hat{A}^\dagger\hat{a}_{S_j}\right)\ket{\psi_{N-1}}
    \\&=
\hat{a}_{I_j}^\dagger\ket{\psi_{N-1}}  +\hat{A}^\dagger\hat{a}_{S_j} \ket{\psi_{N-1}}  
    \\&=
 \hat{a}_{I_j}^\dagger\ket{\psi_{N-1}}  +\frac{\sqrt{N-1}}{\sqrt{N+M-2}} \hat{A}^\dagger \hat{a}_{I_J}^\dagger \ket{\psi_{N-2}} \qquad \text{use inductive hypothesis}
    \\&=
\hat{a}_{I_j}^\dagger\ket{\psi_{N-1}}  + \frac{\sqrt{N-1}}{\sqrt{N+M-2}} \hat{a}_{I_J}^\dagger \hat{A}^\dagger \ket{\psi_{N-2}}  \qquad \text{the idler operators commute with } \hat{A}^\dagger 
    \\&=
 \hat{a}_{I_j}^\dagger\ket{\psi_{N-1}}  +\sqrt{N-1}\sqrt{N+M-2}\frac{\sqrt{N-1}}{\sqrt{N+M-2}}\hat{a}_{I_J}^\dagger \ket{\psi_{N-1}} 
 \\
   \sqrt{N}\sqrt{N+M-1}\hat{a}_{S_j}\ket{\psi_N} &= \left(1+(N-1)\right) \hat{a}_{I_j}^\dagger\ket{\psi_{N-1}} 
\\
\hat{a}_{S_j}\ket{\psi_N} &= \frac{\sqrt{N}}{\sqrt{N+M-1}}\hat{a}_{I_j}^\dagger\ket{\psi_{N-1}} 
 \end{align}

 }

\subsection{Constructing the Projector}
The state $\ket{\tilde{\phi}_{\mathbf{N_A}}}$ is:
\begin{align}
    \ket{\tilde{\phi}_{\mathbf{N_A}}}&= \sqrt{\eta}^{N-N_A}\sqrt{1-\eta}^{N_A}\prod_{i=1}^M\frac{1}{\sqrt{N_{A_i}!}} \left(\hat{a}_{S_i}\right)^{{N_{A_i}}} \ket{\psi_N}_{I,S}
    \\&=
    {N+M-1 \choose M-1}^{-\frac{1}{2}}\sum_{\mathbf{N_S}} \sqrt{\eta}^{N_s-N_A}\sqrt{1-\eta}^{N_A}\left(\mathbf{\hat{a}_S}\right)^{\mathbf{N_A}} \ket{\mathbf{N_S},\mathbf{N_S}}
    \\&=
    \frac{1}{\sqrt{\mathbf{N_A}!}} \sqrt{\eta}^{N-N_A}\sqrt{1-\eta}^{N_A}\left(\mathbf{\hat{a}_S}\right)^{\mathbf{N_A}}\ket{\psi_N}
    \\ &=
    \frac{1}{\sqrt{\mathbf{N_A}!}} \sqrt{\eta}^{N-N_A}\sqrt{1-\eta}^{N_A}\left(\prod_{n=N-N_A+1}^N\frac{\sqrt{n}}{\sqrt{n+M-1}}\right)\left(\mathbf{\hat{a}_I}^\dagger \right)^{\mathbf{N_A}}\ket{\psi_{N-N_A}}
    \\ &=
    \frac{1}{\sqrt{\mathbf{N_A}!}} \sqrt{\eta}^{N-N_A}\sqrt{1-\eta}^{N_A}\sqrt{\frac{N!}{(N-N_A)!} \frac{(N-N_A+M-1)!}{(N+M-1)!}}\left(\mathbf{\hat{a}_I}^\dagger \right)^{\mathbf{N_A}}\ket{\psi_{N-N_A}}
    \\ &=
    \frac{1}{\sqrt{\mathbf{N_A}!}} \sqrt{\eta}^{N-N_A}\sqrt{1-\eta}^{N_A}\sqrt{{N-N_A+M-1 \choose N-N_A} {N+M-1 \choose N}^{-1}}\left(\mathbf{\hat{a}_I}^\dagger \right)^{\mathbf{N_A}}\ket{\psi_{N-N_A}}   
    \\ &=
        \frac{1}{\sqrt{\mathbf{N_A}!}} \sqrt{\eta}^{N-N_A}\sqrt{1-\eta}^{N_A}\sqrt{{N+M-1 \choose N}^{-1}}\left(\mathbf{\hat{a}_I}^\dagger \right)^{\mathbf{N_A}} \sum_{|\mathbf{N_S}|=N-N_A}\ket{\mathbf{N_S},\mathbf{N_S}}
    \\&=
    \frac{1}{\sqrt{\mathbf{N_A}!}} \sqrt{\eta}^{N-N_A}\sqrt{1-\eta}^{N_A}\sqrt{{N+M-1 \choose N}^{-1}}\sum_{|\mathbf{N_S}|=N-N_A}\frac{\sqrt{(\mathbf{N_S}+\mathbf{N_A})!}}{\sqrt{\mathbf{N_S}!}} \ket{\mathbf{N_S}+\mathbf{N_A},\mathbf{N_S}}
\end{align}
Before we construct the projector, we need to compute the norm and inner products. The norm is:
 {\tiny 
 \begin{align}
    \bra{\tilde{\phi}_{\mathbf{N_A}}}\ket{\tilde{\phi}_{\mathbf{N_A}}}
    &=
    \frac{1}{\mathbf{N_A}!} \eta^{N-N_A}(1-\eta)^{N_A}{N+M-1 \choose N}^{-1}\sum_{\substack{|\mathbf{N_S}|=N-N_A\\|\mathbf{N'_S}|=N-N_A}}\frac{\sqrt{(\mathbf{N_S}+\mathbf{N_A})!}}{\sqrt{\mathbf{N_S}!}}\frac{\sqrt{(\mathbf{N'_S}+\mathbf{N_A})!}}{\sqrt{\mathbf{N'_S}!}}\bra{\mathbf{N'_S}+\mathbf{N_A},\mathbf{N'_S}}\ket{\mathbf{N_S}+\mathbf{N_A},\mathbf{N_S}}
    \\&=
    \frac{1}{\mathbf{N_A}!} \eta^{N-N_A}(1-\eta)^{N_A} {N+M-1 \choose N}^{-1}\sum_{|\mathbf{N_S}|=N-N_A}\frac{\sqrt{(\mathbf{N_S}+\mathbf{N_A})!}}{\sqrt{\mathbf{N_S}!}}\frac{\sqrt{(\mathbf{N_S}+\mathbf{N_A})!}}{\sqrt{\mathbf{N_S}!}}\bra{\mathbf{N_S}+\mathbf{N_A}}\ket{\mathbf{N_S}+\mathbf{N_A}}
    \\&=
     \eta^{N-N_A}(1-\eta)^{N_A}{N+M-1 \choose N}^{-1}\sum_{|\mathbf{N_S}|=N-N_A}\frac{(\mathbf{N_S}+\mathbf{N_A})!}{\mathbf{N_S}!\mathbf{N_A}!}
\end{align}
 }
Thus the normalized version is:
\begin{align}
    \ket{\phi_{\mathbf{N_A}}} 
    &=
    \frac{1}{\sqrt{\braket{\tilde{\phi}_{\mathbf{N_A}}}}} \ket{\tilde{\phi}_{\mathbf{N_A}}}
    \\&=
    \frac{1}{\sqrt{\mathbf{N_A}!}} \sqrt{{N-N_A+M-1 \choose N-N_A}}\left(\sum_{|\mathbf{N_S}|=N-N_A}\frac{(\mathbf{N_S}+\mathbf{N_A})!}{\mathbf{N_S}!\mathbf{N_A}!} \right)^{-\frac{1}{2}} \left(\mathbf{\hat{a}_I}^\dagger \right)^{\mathbf{N_A}}\ket{\psi_{N-N_A}}
\end{align}

Now for the mutual inner product:
 \begin{align}
     \bra{\tilde{\phi}_{\mathbf{N_A}}}\ket{\tilde{\phi}_{\mathbf{N'_A}}}
     &\propto
     \sum_{\substack{|\mathbf{N_S}|=N-N_A\\|\mathbf{N'_S}|=N-N'_A}}\frac{\sqrt{(\mathbf{N_S}+\mathbf{N_A})!}}{\sqrt{\mathbf{N_S}!}}\frac{\sqrt{(\mathbf{N'_S}+\mathbf{N'_A})!}}{\sqrt{\mathbf{N'_S}!}}\bra{\mathbf{N'_S}+\mathbf{N'_A},\mathbf{N'_S}}\ket{\mathbf{N_S}+\mathbf{N_A},\mathbf{N_S}}
     \\&=
     \delta_{N_A,N'_A}\sum_{|\mathbf{N_S}|=N-N_A}\frac{\sqrt{(\mathbf{N_S}+\mathbf{N_A})!}}{\sqrt{\mathbf{N_S}!}}\frac{\sqrt{(\mathbf{N_S}+\mathbf{N'_A})!}}{\sqrt{\mathbf{N_S}!}}\bra{\mathbf{N_S}+\mathbf{N'_A}}\ket{\mathbf{N_S}+\mathbf{N_A}}
     \\&=\delta_{\mathbf{N_A},\mathbf{N'_A}}
 \end{align}

Therefore the states $\ket{\phi_{\mathbf{N_A}}}$ are mutually orthonormal and projector is:
\begin{align}
    \hat{P}&=\sum_{|\mathbf{N_A} |\leq{N-1}}\ket{\phi_{\mathbf{N_A}}}\bra{\phi_{\mathbf{N_A}}}
\end{align}

\subsection{False Alarm Probability}
\begin{align}
    &P_{FA}\\ 
    &=
    \Tr\left[ \hat{P} \rho_{abs}\right]
    \\&=
    \Tr\left[ \left( \sum_{|\mathbf{N_A}| \leq{N-1}} \ket{\phi_{\mathbf{N_A}}}\bra{\phi_{\mathbf{N_A}}}\right) \left({N+M-1 \choose M-1}^{-1} \sum_{|\mathbf{N_S}|=N}\sum_{\mathbf{N_B}} p_{\mathbf{N_B}} \ket{\mathbf{N_S},\mathbf{N_B}}\bra{\mathbf{N_S},\mathbf{N_B}} \right)\right]
    \\&=
     {N+M-1 \choose M-1}^{-1}\sum_{|\mathbf{N_A}| \leq{N-1}} \sum_{|\mathbf{N_S}|=N}\sum_{\mathbf{N_B}}  p_{\mathbf{N_B}} \left|\bra{\mathbf{N_S},\mathbf{N_B}} \ket{\phi_{\mathbf{N_A}}}\right|^2
    \\&=
   {N+M-1 \choose M-1}^{-1}\sum_{|\mathbf{N_A}| \leq{N-1}} \sum_{|\mathbf{N_S}|=N}\sum_{\mathbf{N_B}}  p_{\mathbf{N_B}}
   \nonumber \\& \qquad\qquad\qquad\qquad
   \left|\bra{\mathbf{N_S},\mathbf{N_B}} 
   \frac{1}{\sqrt{\mathbf{N_A}!}} \sqrt{{N-N_A+M-1 \choose N-N_A}} \left(\sum_{|\mathbf{N''_S}|=N-N_A}\frac{(\mathbf{N''_S}+\mathbf{N_A})!}{\mathbf{N''_S}!\mathbf{N_A}!} \right)^{-\frac{1}{2}} \left(\mathbf{\hat{a}_I}^\dagger \right)^{\mathbf{N_A}}\ket{\psi_{N-N_A}} \right|^2
    \\&=
    {N+M-1 \choose M-1}^{-1}\sum_{|\mathbf{N_A}| \leq{N-1}} \sum_{|\mathbf{N_S}|=N}\sum_{\mathbf{N_B}}  \frac{p_{\mathbf{N_B}}}{\mathbf{N_A}!} \left(\sum_{|\mathbf{N''_S}|=N-N_A}\frac{(\mathbf{N''_S}+\mathbf{N_A})!}{\mathbf{N''_S}!\mathbf{N_A}!} \right)^{-1}
    \left|\bra{\mathbf{N_S},\mathbf{N_B}} \left(\mathbf{\hat{a}_I}^\dagger \right)^{\mathbf{N_A}}\sum_{|\mathbf{N'_S}|=N-N_A}\ket{\mathbf{N'_S},\mathbf{N'_S}}\right|^2
    \\&=
    {N+M-1 \choose M-1}^{-1}\sum_{|\mathbf{N_A}| \leq{N-1}} \sum_{|\mathbf{N_S}|=N}\sum_{\mathbf{N_B}}  \frac{p_{\mathbf{N_B}}}{\mathbf{N_A}!} \left(\sum_{|\mathbf{N''_S}|=N-N_A}\frac{(\mathbf{N''_S}+\mathbf{N_A})!}{\mathbf{N''_S}!\mathbf{N_A}!} \right)^{-1} 
    \nonumber \\& \qquad\qquad\qquad\qquad\qquad\qquad\qquad\qquad
    \left|\sum_{|\mathbf{N'_S}|=N-N_A}\frac{\sqrt{(\mathbf{N'_S}+\mathbf{N_A})!}}{\sqrt{\mathbf{N'_S}!}}\bra{\mathbf{N_S},\mathbf{N_B}}  \ket{\mathbf{N'_S}+\mathbf{N_A},\mathbf{N'_S}}\right|^2
    \\&=
    {N+M-1 \choose M-1}^{-1}\sum_{|\mathbf{N_A}| \leq {N-1}} \sum_{|\mathbf{N_S}|=N}\sum_{|\mathbf{N_B}|=N-N_A} \frac{p_{\mathbf{N_B}}}{\mathbf{N_A}!} \left(\sum_{|\mathbf{N''_S}|=N-N_A}\frac{(\mathbf{N''_S}+\mathbf{N_A})!}{\mathbf{N''_S}!\mathbf{N_A}!} \right)^{-1} 
    \left|\frac{\sqrt{(\mathbf{N_B}+\mathbf{N_A})!}}{\sqrt{\mathbf{N_B}!}}\bra{\mathbf{N_S}}  \ket{\mathbf{N_B}+\mathbf{N_A}}\right|^2
    \\&=
    {N+M-1 \choose M-1}^{-1}\sum_{|\mathbf{N_A}| \leq {N-1}}  \tilde{p}_{(N-N_A)} \left(\sum_{|\mathbf{N''_S}|=N-N_A}\frac{(\mathbf{N''_S}+\mathbf{N_A})!}{\mathbf{N''_S}!\mathbf{N_A}!} \right)^{-1} \sum_{|\mathbf{N_B}|=N-N_A} \frac{(\mathbf{N_B}+\mathbf{N_A})!}{\mathbf{N_A}!\mathbf{N_B}!}
    \\&=
    {N+M-1 \choose M-1}^{-1}\sum_{N_A=0}^{N-1} {N-N_A+M-1 \choose N-N_A}  \tilde{p}_{(N-N_A)}
    \\&=
    \tilde{p}_{1} \frac{N}{N+M-1} + \tilde{p}_{2} \frac{N(N-1)}{(N+M-1)(N+M-2)} + ... + \tilde{p}_{N} {M+N-1 \choose N}^{(-1)}
\end{align}

\subsection{Probability of Missed Detection}
\begin{align}
1-P_{MD} 
&=
    Tr\left[\hat{P} \rho_{pres} \right]
    \\&=
    Tr\left[\left(\sum_{|\mathbf{N_A} |\leq{N-1}} \ket{\phi_{\mathbf{N_A}}}\bra{\phi_{\mathbf{N_A}}} \right) \left(\sum_{|\mathbf{N'_A}|\leq N} \ket{\tilde{\phi}_{\mathbf{N'_A}}}\bra{\tilde{\phi}_{\mathbf{N'_A}}}\right)\right]
    \\&=
    \sum_{|\mathbf{N_A} |\leq{N-1}}\sum_{|\mathbf{N'_A}|\leq N}  \frac{1}{\braket{\tilde{\phi}_{\mathbf{N_A}}}}Tr\left[ \ket{\tilde{\phi}_{\mathbf{N_A}}}\bra{\tilde{\phi}_{\mathbf{N_A}}} \ket{\tilde{\phi}_{\mathbf{N'_A}}}\bra{\tilde{\phi}_{\mathbf{N'_A}}} \right]
    \\&=
    \sum_{|\mathbf{N_A} |\leq{N-1}}\sum_{|\mathbf{N'_A}|\leq N}  \left| \bra{\tilde{\phi}_{\mathbf{N_A}}} \ket{\tilde{\phi}_{\mathbf{N'_A}}} \right|
    \\&=
    \sum_{|\mathbf{N_A} |\leq{N-1}} \left| \bra{\tilde{\phi}_{\mathbf{N_A}}} \ket{\tilde{\phi}_{\mathbf{N_A}}} \right|
    \\&= 
    1-\sum_{|\mathbf{N_A}| =N} \left| \bra{\tilde{\phi}_{\mathbf{N_A}}} \ket{\tilde{\phi}_{\mathbf{N_A}}} \right|
    \\&= 
    1-\sum_{|\mathbf{N_A}| =N}      \eta^{N-N}(1-\eta)^{N} {N+M-1 \choose N}^{-1}\sum_{|\mathbf{N_S}|=N-N}\frac{(\mathbf{N_S}+\mathbf{N_A})!}{\mathbf{N_S}!\mathbf{N_A}!}
      \\&=  
    1-\sum_{|\mathbf{N_A}| =N}      \eta^{0}(1-\eta)^{N} {N+M-1 \choose N}^{-1}\frac{(\mathbf{0}+\mathbf{N_A})!}{\mathbf{0}!\mathbf{N_A}!}
    \\&=1-(1-\eta)^{N}{N+M-1 \choose N}^{-1}\sum_{|\mathbf{N_A}| =N} 1      
    \\&=
    1-(1-\eta)^{N}
\end{align}

\end{document}